\documentclass[prl,aps,floats,twocolumn]{revtex4}

\usepackage{graphics}
\usepackage{amssymb}
\usepackage{color}

\def\nle{\ \raise.3ex\hbox{$<$}\kern-0.8em\lower.7ex\hbox{$\sim$}\ }
\def\nge{\ \raise.3ex\hbox{$>$}\kern-0.8em\lower.7ex\hbox{$\sim$}\ }

\begin{document}

\title{N\'eel Temperature of Quasi-Low-Dimensional Heisenberg 
Antiferromagnets}

\author{C.~Yasuda$^1$, S.~Todo$^2$, K.~Hukushima$^3$,
F.~Alet$^{4,5,6}$, M.~Keller$^4$, M.~Troyer$^{4,5}$, and 
H.~Takayama$^7$}

\affiliation{$^1$ Department of Physics, Aoyama Gakuin University,
Sagamihara 229-8558, Japan}
\affiliation{$^2$ Department of Applied Physics, University of Tokyo, 
Tokyo 113-8656, Japan}
\affiliation{$^3$ Department of Basic Science, University of Tokyo, 
Tokyo
153-8902, Japan}
\affiliation{$^4$ Theoretische Physik, Eidgen\"ossische Technische 
Hochschule,
    CH-8093 Z\"urich, Switzerland}
\affiliation{$^5$ Computational Laboratory, Eidgen\"ossische Technische 
Hochschule,
    CH-8092 Z\"urich, Switzerland}
\affiliation{$^6$ Service de Physique Th\'eorique, CEA Saclay, 91191 Gif
sur Yvette, France}
\affiliation{$^7$ Institute for Solid State Physics, University of 
Tokyo, Kashiwa 277-8581, Japan}

\date{\today}
%\date{December 03, 2003}

\begin{abstract}
The N\'eel temperature, $T_{\rm N}$, of quasi-one- and 
quasi-two-dimensional
antiferromagnetic Heisenberg models on a cubic lattice is calculated
by Monte Carlo simulations as a function of inter-chain (inter-layer)
to intra-chain (intra-layer) coupling $J'/J$ down to $J'/J\simeq 
10^{-3}$.
We find that $T_{\rm N}$ obeys a modified random-phase
approximation-like relation for small $J'/J$ with an effective universal
renormalized coordination number, independent of the size of the spin.
Empirical formulae describing $T_{\rm N}$ for a wide range of $J'$
and useful for the analysis of experimental measurements are presented.
\end{abstract}


\maketitle

While genuinely one-dimensional (1D) and two-di\-men\-sion\-al (2D)
antiferromagnetic Heisenberg (AFH) models cannot display
long-range order (LRO) except at zero temperature~\cite{Mermin}, weak
inter-chain or inter-layer couplings, $J'$, which always exist in real
materials, lead to a finite N\'eel temperature $T_{\rm N}$. So far, the
$J'$-dependence of $T_{\rm N}$ was calculated by exactly treating 
effects
of the strong interaction $J$ in the 1D or 2D system, but using
mean-field approximations for the inter-chain and inter-layer coupling
$J'$~\cite{Scalapino}. Recently, more advanced theories of the latter
effects have been proposed for quasi-1D (Q1D)~\cite{Irkhin,Bocquet} and
quasi-2D (Q2D)~\cite{Irkhin_Q2D} systems, and the results have been
compared with the experimental observations on Q1D antiferromagnets, 
e.g.,
Sr$_2$CuO$_3$~\cite{Keren}, and Q2D antiferromagnets, e.g.,
La$_2$CuO$_4$~\cite{Keimer}.
In view of the importance of experimentally
well-studied Q2D antiferromagnets as undoped parent compounds of the
high-temperature superconductors, accurate and unbiased numerical
results for Q1D and Q2D AFH models are strongly desired.
In a recent work along this line, Sengupta {\it et al}.~\cite{Sengupta}
have demonstrated peculiar temperature
dependences of the specific heat in the quantum Q2D AFH model.

Here we calculate the N\'eel temperature $T_{\rm N}$ as a function of
$J'$ in fully three-dimensional (3D) classical and quantum Monte
Carlo (MC) simulations of coupled-chains and coupled-layers.
Our MC results on the quantum spin-$S$ and classical $S=\infty$  AFH
models are analyzed by a modified random-phase
approximation (RPA) with a renormalized coordination number defined by
\begin{equation}
     \label{delta-chi}
     \zeta(J') \equiv \frac{1}{J'\chi_{\rm s}(T_{\rm N}(J'))} \ ,
\end{equation}
where $\chi_{\rm s}(T)$ is the staggered susceptibility of the 1D or
2D model at temperature $T$.

In a simple RPA calculation~\cite{Scalapino}, this quantity is just the
coordination number $z_d$ in the inter-chain or inter-layer directions:
$z_1=4$ and $z_2=2$ for the Q1D and Q2D systems, respectively. Our main
result is that $\zeta(J')$ evaluated by Eq.~(\ref{delta-chi}) with our
numerically obtained $T_{\rm N}(J')$ and $\chi_{\rm s}(T)$ becomes
constant
\begin{equation}
     \label{IK-formula}
     \zeta(J') \approx \zeta_d = k_d z_d
\end{equation}
for $J' < J'_{\rm c} \simeq 0.1J$, with the constants  $k_1 = 0.695$ and
$k_2=0.65$. These constants $k_d$ differ from the simple RPA result
$k_d=1$, but the value of $k_1$ is consistent with the modified
self-consistent RPA theory for the quantum Q1D (q-Q1D) model of Irkhin
and Katanin (IK)~\cite{Irkhin}. Furthermore we find, that, within our
numerical accuracy, the value of $k_d$ is the same for the $S=1/2$,
$S=1$, $S=3/2$ and $S=\infty$, and we  conjecture that $k_d$ is
universal and independent of the spin $S$ for small $J'/J$.

We also propose empirical formulae for $T_{\rm N}(J')$ for all values of
$J'$ examined in the present work up to $J'=J$ where corrections to the
modified RPA are significant quantitatively. These formulae  are useful
in analyzing experimental results on infinite-layer antiferromagnets 
such
as Ca$_{0.85}$Sr$_{0.15}$CuO$_2$~\cite{infinite}, (5CAP)$_2$CuBr$_4$ and
(5MAP)$_2$CuBr$_4$~\cite{Woodward}, where they allow to determine the
strength of the inter-chain or inter-layer coupling $J'$ from
experimental measurements of $T_{\rm N}$.

{\it Model and numerical methods.}---
The Hamiltonian of the Q1D and Q2D AFH models is defined on an
anisotropic simple cubic lattice:
\begin{eqnarray}
  \label{Ham}
   {\cal H} &=& \sum_{i,j,k} (J_x {\bf S}_{i,j,k} \cdot {\bf S}_{i+1,j,k}
            + J_y {\bf S}_{i,j,k} \cdot {\bf S}_{i,j+1,k} \nonumber \\
            &+& J_z {\bf S}_{i,j,k} \cdot {\bf S}_{i,j,k+1}) \ ,
\end{eqnarray}
where the summation $\sum_{i,j,k}$ runs over all the lattice sites
on an $L_x \times L_y \times L_z$ cubic lattice
and ${\bf S}_{i,j,k}$ is the spin operator at site ($i,j,k$).
We put $J_x=J_y=J'$ and $J_z=J$ for the Q1D model and $J_x=J_y=J$ and
$J_z=J'$ for the Q2D model with $J >0$ and $0 \le J' \le J$.
For comparison, we also examine the classical limit $S=\infty$ of
Eq.~(\ref{Ham}). Note that $JS(S+1)$ sets the energy scale in the
classical model.  In practice, we simulate a system of unit vectors,
which is equivalent to fixing $JS(S+1)$ to unity.

The MC simulations have been performed using the
continuous-imaginary-time loop algorithm~\cite{QMC} for the quantum
model (QMC) and the Wolff cluster algorithm~\cite{Wolf} for the
classical model (CMC). The AF correlation length $\xi_\alpha$ in each of
the directions $\alpha\ (=x, y, z)$ are evaluated by the second-moment
method~\cite{Cooper} on lattices whose aspect ratio is chosen such
that $\xi_\alpha/L_\alpha$ does not depend on $\alpha$ in the vicinity
of $T_{\rm N}$. Explicitly, for the $S=1/2$ q-Q1D systems with
$J'/J=$0.01, 0.05, 0.1, and 0.5, we set $L_z/L_x=36$ ($L_z \le 504$), 12
(384), 4 (200), and 2 (128), respectively, while for the $S=1/2$ quantum
Q2D (q-Q2D) systems with $J'/J=0.001$, 0.005, 0.01, and $\ge 0.02$, we
set $L_x/L_z=48$ ($L_x \le 288$), 20 (240), 16 (192), and 1 (80),
respectively. Then we determine $T_{\rm N}$ from finite-size scaling,
looking for the best data collapse of $\xi_\alpha / L_\alpha$ plotted
versus $(T-T_{\rm N})L^{1/\nu}$ for different system sizes. We have
fixed the exponent $\nu=0.71$~\cite{Chen} to the value of the 3D
classical Heisenberg universality class.
The values of $T_{\rm N}$ obtained for the   $S=1/2$ q-Q1D and 
classical Q1D
(c-Q1D) systems, and the  $S=1/2$ q-Q2D and classical Q2D (c-Q2D) 
systems are
listed in Table~\ref{table-tn}.

\begin{table}[t]
  \caption{N\'eel temperatures of the $S=1/2$ q-Q1D, c-Q1D,  $S=1/2$
  q-Q2D, and c-Q2D AFH models, normalized by $JS(S+1)$. The result for
  the classical system with $J'/J=1$ is taken from 
Ref.~\protect{\cite{Chen}}.}

\begin{tabular}{l||l|l|l|l} \hline\hline
$J'/J$ & \multicolumn{4}{c}{$T_{\rm N}/JS(S+1)$} \\ \hline
        & q-Q1D & c-Q1D & q-Q2D & c-Q2D \\ \hline
     1     & 1.2589(1)   & 1.4429(1)& 1.2589(1)  & 1.4429(1) \\
     0.5   & 0.78997(8)  & 0.9317(1) & 1.0050(4)  & 1.1733(1) \\
     0.1   & 0.22555(3)  & 0.39551(8) & 0.6553(4)  & 0.8526(1) \\
     0.05  & 0.12171(5)  & 0.28377(4) & 0.5689(2)  & 0.7797(1) \\
     0.02  & 0.05258(1)  & 0.18361(3) & 0.48463(8) & 0.7115(1) \\
     0.01  & 0.02768(1)  & 0.13157(2) & 0.43515(6) & 0.6731(2) \\
     0.005 &             & 0.09393(2) & 0.39513(4) & 0.6419(2) \\
     0.001 &             & 0.042547(6) & 0.32571(8) & 0.5858(4) \\ 
\hline\hline
\end{tabular}
  \label{table-tn}
\end{table}%

{\it Q1D systems.}---
The classical 1D (c-1D) model shows LRO at $T=0$, while the ground
state of the $S=1/2$ and $S=3/2$ quantum 1D (q-1D) model is gapless
and has no LRO~\cite{Bethe}. Correspondingly, the staggered
susceptibility $\chi_{\rm s}$ for the classical model, given exactly
by~\cite{Nakamura}
\begin{equation}
  \chi_{\rm s}(T) = 
\frac{x}{3J}\frac{1+1/\tanh{x}-1/x}{1-1/\tanh{x}+1/x} \
\label{c1d-exact}
\end{equation}
with $x=JS(S+1)/T$, diverges as $T^{-2}$ in the limit
$T \rightarrow 0$, while the one for the $S=1/2$ model, asymptotically
given by~\cite{Barzykin}
\begin{equation}
   \chi_{\rm s}(T) \simeq \frac{c_1}{T}
\sqrt{\ln{(\frac{\Lambda J}{T})}+\frac{1}{2}\ln{\ln{(\frac{\Lambda 
J}{T})}}} \ ,
  \label{1d-ssus}
\end{equation}
exhibits only a $1/T$ divergence with logarithmic corrections. Here we
note that the quantitative accuracy of this expression is limited to a
very low temperature range. In fact Eq.~(\ref{1d-ssus}) with the
constants $c_1$ and $\Lambda$ derived
field-theoretically~\cite{Barzykin} does not fit well to $\chi_{\rm s}$
calculated numerically at $T \ge 0.003J$. This indicates the limits of
applicability of analytical results and that one has to use instead
numerical data in this temperature range.

\begin{figure}[t]
  \centerline{\resizebox{0.45\textwidth}{!}{\includegraphics{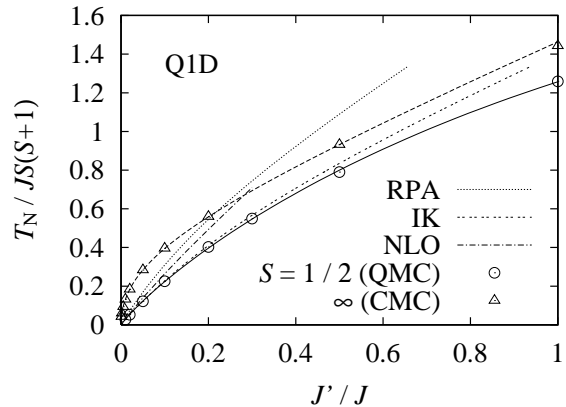}}}
  \vspace*{-.5em}
  \caption{$J'$-dependence of $T_{\rm N}/JS(S+1)$ for the Q1D  systems.
The open symbols denote our MC results. The error bar
  of each point is much smaller than the symbol size.  The dashed curve
  passing through the $S=\infty$ data is obtained by
  Eq.~(\ref{IK-formula}) with $k_1$=0.7 and
  Eq.~(\ref{c1d-exact}) for $\chi_{\rm s}$.  The solid curve denotes the
  proposed empirical formula of Eq.~(\ref{q1d-tn-j-2}). The other curves
  are results of various approximations discussed in the text.}
  \label{q1d-tn-paper}
\end{figure}
Due to the different functional forms of the quantum and classical
susceptibilities, we observe in Fig.~\ref{q1d-tn-paper} that, at small
$J'/J$, $T_{\rm N}(J') \propto \sqrt{J'/J}$ for the classical
model, while $T_{\rm N}(J') \propto J'/J$ with logarithmic corrections
for the quantum model. Comparing the RPA result (Eq.~(\ref{IK-formula})
with $k_1=1$) with the modified RPA one (Eq.~(\ref{IK-formula}) with
$k_1 \simeq 0.70$, denoted by IK), one can easily see that the latter
describes $T_{\rm N}(J')$ much better and is a fairly good description
of $T_{\rm N}(J')$ in the range $J'/J \lesssim 0.3$. Comparing our
results to the next leading order finite-temperature perturbation
theory~\cite{Bocquet} (NLO in Fig.~\ref{q1d-tn-paper}) which is based on
Eq.~(\ref{1d-ssus}), however, we do not find good agreement, because, as
pointed out above, Eq.~(\ref{1d-ssus}) is inappropriate in the
considered temperature range.

\begin{figure}[t]
  \centerline{\resizebox{0.45\textwidth}{!}{\includegraphics{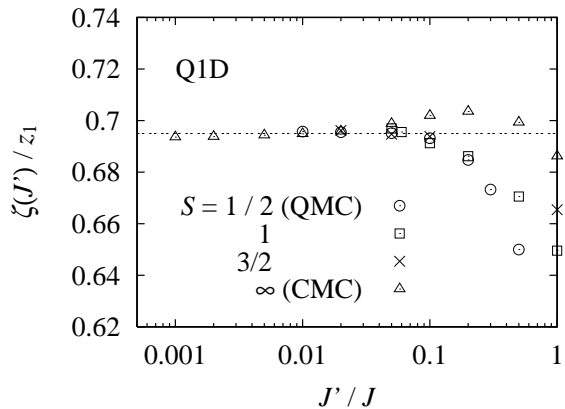}}}
  \vspace*{-.5em}
  \caption{$J'$-dependence of $\zeta(J')/z_1$ for the Q1D systems. In 
all cases
  $\zeta(J')/z_1$ approaches a constant ($\simeq 0.695$), denoted by
  the dotted line, at small $J'/J$. The error bar of each point is
  smaller than the symbol size.
}
  \label{q1d-dssus-paper}
\end{figure}

The agreement with the modified RPA theory is directly shown in
Fig.~\ref{q1d-dssus-paper} where the $J'$-dependence of $\zeta(J')$ in
Eq.~(\ref{delta-chi}) is shown. The $\chi_{\rm s}(T_{\rm N}(J'))$ are
obtained from QMC simulations interpolated near $T=T_{\rm N}$ for the
$S=1/2$ model and from Eq.~(\ref {c1d-exact}) for the
$S=\infty$ model. For $J'/J\le 0.1$ we reach Eq.~(\ref{IK-formula}) with
$k_1 \simeq 0.695$ for the $S=1/2$ model
as well as for the classical limit $S=\infty$ model and
conclude that, within the numerical accuracy of our simulation,
the modified RPA with $J'$-independent $\zeta(J')$ is an appropriate
quantitative description of the models with $J'/J$ in this range.

Interestingly the result mentioned above seems to hold
for quantum models with other values of $S$. As also shown in
Fig.~\ref{q1d-dssus-paper}, within our numerical accuracy, this is well
confirmed for the $S=3/2$ model with $J'/J \ge 0.02$. For the $S=1$
case we find agreement in the range $J'/J \ge 0.05$, where $T_N$ is
larger than the Haldane gap~\cite{Haldane} of the isolated chain. Below
this temperature the finite size scaling of the QMC data becomes less
reliable and we cannot draw definitive conclusions. Even if the result
for the $S=1$ model is restricted to this temperature range, the present
result is surprising, given the different behavior of $\chi_{\rm s}(T)$
in the c-1D and q-1D models.

{\it Q2D systems.}---
In both classical 2D (c-2D) and quantum 2D (q-2D) models, AF-LRO appears
at $T=0$, together with an exponential divergence of $\chi_{\rm s}(T)$
at $T \to 0$.  In the c-2D system, $\chi_{\rm s}$ is
proportional to $T^3 \exp(4\pi J/T)$ at low temperatures~\cite{Brezin}
and our numerical results agree with previous
simulations~\cite{Apostolakis}. For the q-2D models, there is a similar
exponential divergence at $T \to 0$.
In the renormalized
classical regime of the non-linear $\sigma$ model~\cite{Chakravarty},
for example, $\chi_{\rm s}(T)$ is written as
\begin{equation}
\chi_{\rm s}(T) J = c_2 T/J~{\rm exp}(4\pi\rho_{\rm s}/T),
  \label{2d-ssus}
\end{equation}
where $\rho_{\rm s}$ is the spin stiffness and $c_2$ a constant.

\begin{figure}[t]
  \centerline{\resizebox{0.45\textwidth}{!}{\includegraphics{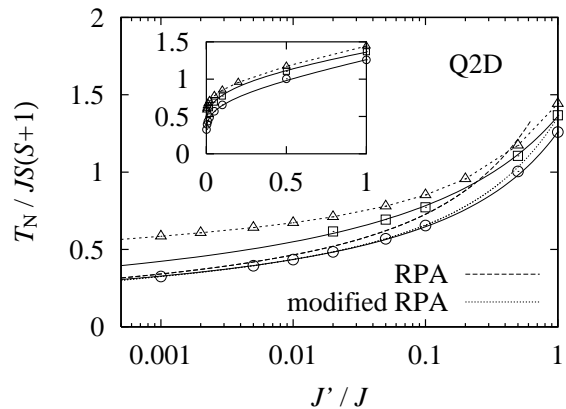}}}
  \vspace*{-.5em}
  \caption{$J'$-dependence of $T_{\rm N}/JS(S+1)$ for the Q2D
  systems. The open symbols denote our numerical results for the
  $S=1/2$ (circles), $S=1$ (squares), and $S=\infty$ (triangles) models.
  The error bar of each point is much smaller than the symbol size. The
  dashed (RPA) and dotted (modified RPA) curves for $S=1/2$ are obtained 
from
  Eqs.~(\ref{delta-chi}) and (\ref{IK-formula}) with $k_2=1$ and
  $k_2=0.65$, respectively.  The solid curves denote the proposed
  empirical formula~(\ref{tn2}), while the curve passing through the
  $S=\infty$ data is simply a guide for the eye. The inset shows the
  same data on a linear scale.}
\label{tn-fit-multi}
\end{figure}

\begin{figure}[t]
  \centerline{\resizebox{0.45\textwidth}{!}{\includegraphics{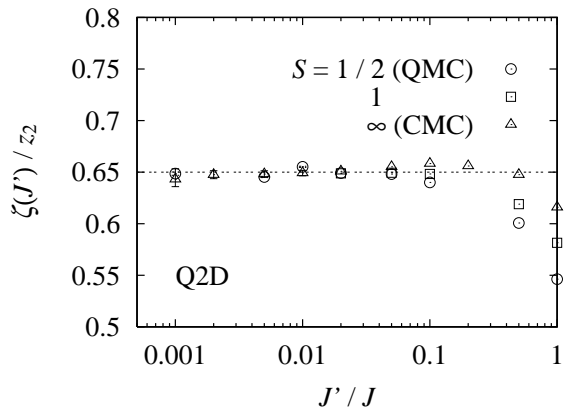}}}
  \vspace*{-.5em}
  \caption{$J'$-dependence of $\zeta(J')/z_2$ for the Q2D systems. In 
all cases
  $\zeta(J')/z_2$ approaches a constant ($\simeq 0.65$), denoted by the
  dotted line, at small $J'/J$.  The error bar of each point is smaller
  than the symbol size unless given explicitly.}
  \label{ssus-tn-paper}
\end{figure}

The $J'$-dependence of $T_{\rm N}$ for the Q2D models is shown in
Fig.~\ref{tn-fit-multi}. We see that $T_{\rm N}(J') 
\propto
-1/\ln{(J'/J)}$ at small $J'/J$ in the $S=1/2$,  $S=1$ and $S=\infty$
models due to the similar exponential forms of
$\chi_{\rm s}$ at $T\rightarrow 0$ of the
classical and quantum models. Figure~\ref{ssus-tn-paper} shows that
again for $J'/J \nle 0.05$ the values of $\zeta(J')$ are  universal for
the quantum and the classical models: $k_2=0.65$ in
Eq.~(\ref{IK-formula}) independent of the spin size $S$. 
This confirms the validity of our modified RPA scenario represented by
Eqs.~(\ref{delta-chi}) and (\ref{IK-formula}) also for the Q2D systems.

If we insert
Eq.~(\ref{2d-ssus}) into Eq.~(\ref{delta-chi}) with
$\zeta(J')=\zeta_2$, we obtain the following expression
of $T_{\rm N}$ for $J'/J \ll 1$,
\begin{equation}
  T_{\rm N} = 4\pi\rho_{\rm s}/(b-\ln{(J'/J)}-\ln{(T_{\rm N}/J)}) \ ,
  \label{tn2-low}
\end{equation}
with $b=-{\rm ln}(\zeta_2c_2)$. This result is compatible with that of
the $1/N$-expansion theory of $T_{\rm N}(J')$ due to 
Irkhin {\it et al.}~\cite{Irkhin_Q2D} for the $S=1/2$ model in the same
limit. 
Various estimations of $b$ and $\rho_s$ can be obtained
analytically~\cite{Irkhin_Q2D}
according to the different approximation schemes used.
Unfortunately, we cannot judge which approximation is most relevant
in general since higher order corrections in $T/4\pi\rho_s$ over the
leading asymptotic expression Eq.~(\ref{tn2-low}) are known to be
necessary~\cite{Kim} to reproduce the numerically obtained $\chi_s$
in the temperature range $T/4\pi\rho_s\nge 0.1$ simulated. In fact,
corrections of this type and uncertainty on $T_N$ due to the
different possible estimates of $b$ are comparable.
We expect, on the other hand, that the constancy of
the normalization factor $k_2$, which is found numerically to be within
2\% in $0.001 \nle J'/J \nle 0.05$ and $0.32 \nle T_{\rm N}/JS(S+1) 
\nle 0.57$ (Fig. 4), holds in the limit $J'/J \rightarrow 0$ as well.

{\it Empirical formulae.}---
Finally, we propose empirical formulae for $T_{\rm N}(J')$ based on our
QMC results. % in a wide range of couplings.
For the $S=1/2$ q-Q1D system we propose a modified RPA form based on
Eqs.~(\ref{IK-formula}) and (\ref{1d-ssus}) with a constant $\zeta(J')$,
\begin{equation}
   J' = T_{\rm N}/(4c\sqrt{\ln{(\frac{\lambda J}{T_{\rm 
N}})}+\frac{1}{2}\ln{\ln{(\frac{\lambda J}{T_{\rm N}})}}}) \ ,
  \label{q1d-tn-j-2}
\end{equation}
but with modified values of the constants with $c=0.233$ and
$\lambda=2.6$. These values are chosen to reproduce not $\chi_{\rm
s}(T)$ but $T_{\rm N}(J')$ in a wide range of $J'$. This formula
describes $T_{\rm N}(J')$ very well in the whole range of $J'/J$ as
shown in Fig.~\ref{q1d-tn-paper}, and it can be used to analyze
experimental results, e.g. to obtain $J'/J \simeq 0.0007$ for
Sr$_2$CuO$_3$ from $T_{\rm N}/J \simeq 0.002$~\cite{Keren}.

For the q-Q2D systems, we find that instead of Eq.~(\ref{tn2-low}), the
following simpler expression describes $T_{\rm N}$ better in the range
$0.001 \le J'/J \le 1$ (see Fig.~\ref{tn-fit-multi}):
\begin{equation}
  T_{\rm N} = 4\pi\rho_{\rm s}/(b-\ln{(J'/J)}) \ ,\
  \label{tn2}
\end{equation}
with $\rho_{\rm s}/J=0.183$ and $b=2.43$ for $S=1/2$, and
$\rho_{\rm s}/J=0.68$ and $b=3.12$ for $S=1$. Table~\ref{table-jp}
shows inter-layer couplings $J'$ estimated using this equation for a
number of infinite-layer materials with $S=1/2$.

\begin{table}[t]
\vspace{-5mm}
  \caption{Inter-layer coupling $J'$ estimated by Eq.~(\ref{tn2}) for
  various infinite-layer compounds.  The N\'eel temperatures $T_{\rm
  N}$, the intra-layer couplings $J$, and their ratio estimated by the
  experiments are also listed.}
\begin{tabular}{l|c|c|c|c}
\hline \hline
Compound & $T_{\rm N}$ & $J$ & $T_{\rm N}/J$& $J'/J$ \\ \hline
Ca$_{0.85}$Sr$_{0.15}$CuO$_2$ \cite{infinite} & 537K  & 1535K &
0.35 & 0.016 \\
(5CAP)$_2$CuBr$_4$ \cite{Woodward} & 5.08K & 8.5K & 0.60 & 0.24 \\
(5MAP)$_2$CuBr$_4$ \cite{Woodward} & 3.8K & 6.5K & 0.58 & 0.22 \\
(5CAP)$_2$CuCl$_4$ \cite{Woodward} & 0.74K & 1.14K & 0.64 & 0.33 \\
(5MAP)$_2$CuCl$_4$ \cite{Woodward} & 0.44K & 0.76K & 0.57 & 0.21 \\

\hline \hline
\end{tabular}
\label{table-jp}
\end{table}%

To conclude, we have determined, by high-precision Monte Carlo
simulations, the N\'eel temperatures of quantum and classical Q1D and
Q2D Heisenberg antiferromagnets. Besides finding empirical formulae for
$T_{\rm N}(J')$, we observe that, using numerically accurate values of
the staggered susceptibility a modified RPA with the $J'$-independent
renormalized coordination number $\zeta_d$ succeeds in quantitatively
describing the relation between $T_{\rm N}$ and $J'/J$ for $J'< J'_{\rm
c}$ with $J'_{\rm c} \simeq 0.1J$.

An intriguing result of our simulations is the independence of $\zeta_d$
on the value of the spin $S$, suggesting a universality of corrections
to RPA for $J'\ll J$. Since in this temperature regime the physics of
all these models should be well described by an anisotropic non-linear
$\sigma$-model (NL$\sigma$M) in the renormalized classical regime, we
conjecture universal corrections to RPA also for the NL$\sigma$M.

We acknowledge fruitful discussions with C.P. Landee and M. Matsumoto,
and thank M. Bocquet for useful discussions and for sending his NLO 
results.
Most of the numerical calculations have been performed
on the SGI 2800 at Institute for Solid State Physics, University of
Tokyo. The program is based on `Looper version 2' developed by S.T. and
K. Kato and `PARAPACK version 2' by S.T.  This work is supported by
Grant-in-Aid for Scientific Research Programs (\#12640369, \#15740232)
and 21st COE Program from the Ministry of Education, Science, Sports,
Culture and Technology of Japan, and by the Swiss National Science
foundation.

\end{document}